\magnification=1200
\centerline{HYDRODYNAMIC TURBULENCE AS A PROBLEM}
\centerline{IN NONEQUILIBRIUM STATISTICAL MECHANICS.}
\bigskip\bigskip
\centerline{by David Ruelle\footnote{$\dagger$}{Math. Dept., Rutgers University, and 
IHES, 91440 Bures sur Yvette, France. email: ruelle@ihes.fr}.}
\bigskip\bigskip
{\sl Abstract:}
We reformulate the problem of hydrodynamic turbulence as a heat flow problem.  We obtain thus a prediction
$$	\zeta_p={p\over3}-{1\over\ln\kappa}\ln\Gamma({p\over3}+1)   $$
for the exponents of the structure functions ($\langle|\Delta_rv|^p\rangle=r^{\zeta_p}$).  The meaning of the adjustable parameter $\kappa$ is that when an eddy of size $r$ has decayed to eddies of size $r/\kappa$ their energies have a thermal distribution.  The above formula, with $(\ln\kappa)^{-1}=.32\pm.01$ is compatible with experimental data.  This agreement lends supports to our physically motivated picture of turbulence.
\vfill\eject
	Hydrodynamic turbulence is known to be a chaotic phenomenon [1,2,3].  This means that the time evolution $(f^t)$ of a turbulent fluid system belongs to a much studied class of deterministic dynamics with sensitive dependence on initial conditions [4,5,6].  The statistical properties of turbulence are described by an ergodic invariant state $\rho$ for $(f^t)$ and, since chaotic dynamical systems have (uncountably) many ergodic states, a choice has to be made.  A physically reasonable choice is that of so-called SRB states (see [7,8] and the references given there).
\medskip 
	It is fair to say that the chaotic nature of turbulence has been largely ignored by the turbulence community, and that the choice of an ergodic state to describe the statistical properties of turbulence has been made by ad hoc assumptions (closure assumptions, Gaussianity, multifractal structure).  Indeed, the study of SRB or ``physical'' states for the Navier-Stokes dynamics appears at first impossibly difficult.  Nevertheless we propose here an approach of this sort: we bypass the mathematical problems of SRB states by using our understanding of the physics of a specific dynamical system, namely that corresponding to heat conduction, as seen from the point of view of nonequilibrium statistical mechanics.  Our approach will thus use basic physical ideas, and approximations, rather than ad hoc assumptions.  We shall in this manner obtain a surprisingly coherent view of the fluctuations in turbulence (intermittency).
\medskip
	We shall concern ourselves with incompressible fluids in 3 dimensions, described by the Navier-Stokes equation, but without paying too much attention to the specific form of the dissipative term.  The fluid, with velocity field $v$, will be enclosed in a cube $C_0$ of side $\ell_0$, which we may consider for simplicity to have periodic boundary conditions.  We choose an integer $\kappa>1$ and divide $C_0=C_{01}$ into cubes $C_{ni}$ of side $\ell_n=\ell_0\kappa^{-n}$, with $i=1,\ldots,\kappa^{3n}$, where $n$ is a positive integer.  Let $\phi_{ni}$ be the homothety mapping $C_0$ to $C_{ni}$.  One can choose $2(\kappa^3-1)$ real vector fields $U_\alpha$ on ${\bf R}^3$ with $\int U_\alpha=0$, ${\rm div}\,U_\alpha=0$, and such that if the velocity field $v$ satisfies $\int v=0$, ${\rm div}\,v=0$, there is a unique representation
$$	v=\sum_{n=0}^\infty\sum_{i=1}^{\kappa^{3n}}\sum_{\alpha=1}^{2(\kappa^3-1)}
	c_{ni\alpha}U_\alpha\circ\phi_{ni}^{-1}   $$
with $c_{ni\alpha}\in{\bf R}$.  This means that $v$ has a wavelet decomposition into components (roughly) localized in the cubes $C_{ni}$.
\medskip
	We think now of the standard physical situation where energy is put into the fluid at large spatial wavelength (i.e., small $n$) and dissipated at small spatial wavelength (i.e., large $n$).  Intermediate values of $n$ correspond to the inertial range, where the time evolution should in some sense be Hamiltonian.  Specifically, Arnold [9] has shown how an inviscid flow could be interpreted as geodesic flow on the group of volume preserving diffeomorphisms.  The corresponding Hamiltonian is the kinetic energy of the velocity field.
\medskip
	We may thus think of the time evolution for (the finitely many) coefficients $c_{ni\alpha}$ as Hamiltonian, with external forces acting at low and high $n$.  This is related to the physical concept of eddies as dynamical structures localized in space.  However, instead of a cascade of eddies of smaller and smaller size, we think of a system of coupled Hamiltonian systems which we can label $(n,i)$.  If we assume that the different systems $(n,i)$ are weakly coupled, we can reinterpret the global dynamics as a heat flow from the small $n$, where energy is input, to large $n$ where it is dissipated, i.e., rapidly carried away to structures of molecular size of the fluid.  Note that the multifractal description of eddy cascades [10,11,12] ignores interactions between $(n,i)$, $(n,i')$ except when these eddies are created from a common $(n-1,j)$.  This corresponds to saying that the lateral interaction between the systems $(n,i)$, $(n,i')$ is weak, but this assumption does not appear to be essential in our approach.
\medskip
	There is no hope for an exact study of the dynamics of the coupled systems $(n,i)$, but we can get a first approximation from the Kolmogorov scaling theory of homogeneous turbulence [13].  Since this theory gives unique answers, the problem of selecting an SRB state does not occur here.  According to the Kolmogorov theory, the fluid velocity corresponding to $C_{ni}$ is $v_{ni}\sim(\epsilon\ell_n)^{1/3}$ where $\epsilon$ is the mean dissipation per unit volume, and the kinetic energy corresponding to $C_{ni}$ is
$$	\sim{1\over2}\ell_n^3(\epsilon\ell_n)^{2/3}={\epsilon^{2/3}\over2}\ell_n^{11/3}   $$
(we have put the fluid density equal to 1); the corresponding temperature is
$$	T_n={1\over k}{\epsilon^{2/3}\ell_n^{11/3}\over2(\kappa^3-1)}
={1\over k}{\epsilon^{2/3}\ell_0^{11/3}\kappa^{-11n/3}\over2(\kappa^3-1)}\eqno{(1)}   $$
where $k$ is Boltzmann's constant.  In view of the value of $k$, we see that $T_n$ is huge for small $n$ so that the flow of heat from high temperature to low temperature agrees with the energy cascade from small $\ell$ to large $\ell$ in the fluid.  Notice that the heat resistance $(T_n-T_{n+1})/\epsilon$ is very large, which agrees with a weak coupling between the systems $(n,i)$ for different values of $n$.
\medskip
	We see the situation as follows: a heat flow interpretation of the energy cascade in homogeneous turbulence is possible, using scaling laws, but ignores fluctuations (intermittency).  To understand fluctuations we have to study the fluctuations of the energy flow in the Hamiltonian system of the coupled $(n,i)$.  This is a problem of nonequilibrium statistical mechanics, a problem known to be difficult [14,15,16].  In general one would need the systems $(n,i)$ to be chaotic in some sense (this is physically reasonable for 3-dimensional hydrodynamics) but the Anosov assumptions of [15,16] are unreasonably strong.  In the present situation, a rigorous analysis appears quite out of reach at this time.  An approximate study is however possible, and will give more specific results than earlier multifractal approaches [10,11,12], and there will be a physical justification rather than ad hoc assumptions.
\medskip
	While we have at this time no detailed understanding of heat flow from the point of view of rigorous statistical mechanics, we expect that Fourier's law should hold under normal conditions.  This is however no great help since Kolmogorov's theory yields the precise temperature distribution (1).  As to the ``microscopic'' fluctuations, they are a difficult problem in nonequilibrium [17,18], being different in nature from the well-understood equilibrium fluctuations.  Here we shall use the assumption that the systems $(n,i)$ have weak mutual coupling to justify a Boltzmannian energy distribution for each Hamiltonian system $(n,i)$.
\medskip
	Because of the large temperature gradient, the flow of energy is overwhelmingly from the system $(n,i)$ to the systems $(n+1,j)$.  To study this energy flow we use the conservation of energy and scaling as in the multifractal approaches [11] to write
$$	|v_{ni}|^3/\ell_n=|v_{(n+1)j}|^3/\ell_{n+1}\qquad{\rm or}
	\qquad |v_{(n+1)j}|^3=|v_{ni}|^3\kappa^{-1}\eqno{(2)}   $$
Note that $|v|^3$ is proportional to the kinetic energy ${1\over2}|v|^2$ with a weight 1/ time spent in a certain spatial frequency range.  We interpret then (2) to mean that, given the energy $V_{ni}=|v_{ni}|^3$ in $(n,i)$, the velocity $v=v_{(n+1)j}$ is fluctuating with Boltzmannian distribution
$$	\sim\exp\big(-{|v|^3\over V_{ni}\kappa^{-1}}\big)d^3v   $$
Therefore the energy $V=V_{(n+1)j}$ has the normalized distribution
$$	{1\over V_n\kappa^{-1}}\exp\big(-{V\over V_n\kappa^{-1}}\big)dV\eqno{(3)}   $$
where we write from now on $V_n$ instead if $V_{ni}$, etc.  We view (3) as an approximate, but physically motivated relation, the validity of which will be discussed below.  Note that if we replace $V_n$ by $\tilde V_n=\kappa^nV_n$ we have that $\tilde V=\tilde V_{n+1}$ is distributed according to
$$	{1\over\tilde V_n}\exp(-{\tilde V\over\tilde V_n})d\tilde V   $$
\indent
	We now discuss the structure functions, i.e., the moments
$$	\langle|v_n|^p\rangle=\langle V_n^{p/3}\rangle   $$
for positive integer $p$, and the exponents $\zeta_p$ such that
$$	\langle |v_n|^p\rangle\sim\ell_n^{\zeta_p}\qquad{\rm or}\qquad
	\zeta_p\ln\ell_n\sim\ln\langle V_n^{p/3}\rangle
	=-n\cdot{p\over3}\ln\kappa+\ln\langle\tilde V_n^{p/3}\rangle   $$
We have here
$$	\langle\tilde V_n^{p/3}\rangle
	=\int d\tilde V_1{e^{-\tilde V_1/\tilde V_0}\over\tilde V_0}\int\cdots
	\int d\tilde V_{n-1}{e^{-\tilde V_{n-1}/\tilde V_{n-2}}\over\tilde V_{n-2}}
\int d\tilde V_n{e^{-\tilde V_n/\tilde V_{n-1}}\over\tilde V_{n-1}}\cdot\tilde V_n^{p/3}   $$
$$	\int_0^\infty d\tilde V_n{e^{-\tilde V_n/\tilde V_{n-1}}\over\tilde V_{n-1}}
	\cdot\tilde V_n^{p/3}=\tilde V_{n-1}^{p/3}\int_0^\infty d\xi\,e^{-\xi}\xi^{p/3}
	=\tilde V_{n-1}^{p/3}\Gamma({p\over3}+1)   $$
so that by induction we find
$$	\langle\tilde V_n^{p/3}\rangle=[\Gamma({p\over3}+1)]^n\tilde V_0^{p/3}\qquad,\qquad
	\zeta_p\approx
	{-n{p\over3}\ln\kappa+\ln\langle\tilde V_n^{p/3}\rangle\over-n\ln\kappa}
	\approx{p\over3}-{1\over\ln\kappa}\ln\Gamma({p\over3}+1)   $$
In conclusion we have the (approximate) prediction
$$	\zeta_p={p\over3}-{1\over\ln\kappa}\ln\Gamma({p\over3}+1)\eqno{(4)}   $$
\indent
	Using either the heat propagation or the eddy cascade picture, we see that $\kappa$ should be chosen such that the initial $\tilde V_n$-distribution concentrated on one value for $(n,i)$ thermalizes to values of $\tilde V_{n+1}$ for the systems $(n+1,j)$ distributed according to
$$	{1\over\tilde V_n}e^{-\tilde V_{n+1}/\tilde V_n}d\tilde V_{n+1}   $$
This requires $\kappa$ sufficiently large.  However, if the value of $\kappa$ is too large, several different temperatures will be present among the systems $(n+1,j)$ connected with $(n,j)$, and the $\tilde V_{n+1}$-distribution will not be Boltzmannian.  The picture we have in mind is a situation in $C_{ni}$ which depends on the spatial wavelength: at wavelength of the order of the size of supp$(U_\alpha\circ\phi_{ni}^{-1})$ a single value of the kinetic energy is present; the distribution broadens as the wavelength diminishes and becomes a thermal distribution when it is divided by $\kappa$; at smaller wavelengths there are several patches with different temperatures.  Of course a rigorous justification of this picture is well beyond the power of current mathematical methods.  We can only claim this: $\kappa$ should be such that when an eddy of size $r$ has decayed to eddies of size $r/\kappa$ their energies have a thermal distribution, after which the process can start again.  In the dissipative range the distribution of $V_n$ should be cut off at large $V_n$.  Numerically, one finds that (4) fits the experimental data [19] well ([20] less well) with $1/\log\kappa=.32\pm.01$, i.e., $\kappa$ between 20 and 25.  Note that (4) gives $\zeta_3=1$.
\medskip
	From a physical point of view, one can try the following interpretation: the change of behavior as one passes from large to small wavelength corresponds to what is observed at the onset of turbulence.  We use here the physical fact that transport is much faster in the turbulent than in the laminar regime.  Therefore, when sufficiently small scales are reached and we are in the turbulent regime, thermalization takes place. We may compute the length ratio $\kappa=\ell_n/\ell_{n+1}$ in terms of the Reynolds number $R_c$ for the onset of turbulence.  If $\epsilon$ is the energy dissipation per unit volume and $\nu$ the kinematic viscosity, the Kolmogorov length is $\eta=(\nu^3/\epsilon)^{1/4}$, so that $\nu=(\epsilon\eta^4)^{1/3}$.  The velocity corresponding to the length $\lambda$ is given in the turbulent regime by $v_\lambda=(\epsilon\lambda)^{1/3}$.  Therefore if the onset of turbulence corresponds to $\lambda$, we have
$$	\kappa={\lambda\over\eta}=\big({\lambda^4\epsilon\over\nu^3}\big)^{1/4}
	=\big({\lambda v_\lambda\over\nu}\big)^{3/4}=R_c^{3/4}   $$
The critical Reynolds number $R_c$ is not defined with precision, but the value $R_c=\kappa^{4/3}\approx60$ is not unreasonable.  Clearly the calculation we have made is quite rough, but the exponent $\zeta_p$ should not be very sensitive to details, in particular because $\kappa$ occurs only as its logarithm in (4).  Notice also that the estimate $\kappa=R_c^{3/4}$ is proposed instead of a fundamental calculation which is beyond current possibilities.  Altogether, the agreement of (4) with experiment, with a plausible value of $\kappa$, supports the physical picture of turbulence that we have presented.
\vfill\eject
{\bf References}.
\medskip
[1] D. Ruelle and F. Takens ``On the nature of turbulence.''  Commun. Math. Phys. {\bf 20},167-192(1971) and {\bf 23},343-344(1971).

[2] J.P. Gollub and H.L. Swinney ``Onset of turbulence in a rotating fluid.''  Phys. Rev. Lett. {\bf 35},927-930(1975).

[3] A. Libchaber ``From chaos to turbulence in Benard convection.''  Proc. Roy. Soc. London {\bf A}413,63-69(1987).

[4] E.N. Lorenz ``Deterministic nonperiodic flow.''  J. Atmos. Sci. {\bf 20},130-141(1963).

[5] P. Cvitanovi\'c (editor)  {\it Universality in Chaos}, 2nd ed.  Adam Hilger, Bristol, England, 1989.

[6] Hao Bai-Lin (editor)  {\it Chaos II}.  World Scientific, Singapore, 1990.

[7] L.-S. Young ``What are SRB measures, and which dynamical systems have them?''  J. Statist. Phys. {\bf 108},733-754(2002).

[8] C. Bonatti, L.J. D\'\i az, M. Viana {\it Dynamics beyond uniform hyperbolicity}.  Springer, Berlin, 2005.

[9] V.I. Arnold ``Sur la g\'eom\'etrie diff\'erentielle des groupes de Lie de dimension infinie et ses applications \`a l'hydrodynamique des fluides parfaits.''  Ann. Inst. Fourier {\bf 16},319-361(1966).

[10] G. Parisi and U. Frisch ``On the singularity structure of fully developed turbulence'' in {\it Turbulence and Predictability in Geophysical Fluid Dynamics} (ed. M. Ghil, R. Benzi, and G. Parisi), pp. 84-88.  North-Holland, 1985.

[11] R. Benzi, G. Paladin, G. Parisi, and A. Vulpiani ``On the multifractal nature of fully developed turbulence and chaotic systems.''J. Phys. A {\bf 17},3521-3531(1984).

[12] C. Meneveau and K.R. Sreenivasan ``Simple multifractal cascade model for fully developed turbulence.''  Phys. Rev. Lett. {\bf 59},1424-1427(1987).

[13] A.N. Kolmogorov ``The local structure of turbulence in incompressible viscous fluid for very large Reynolds number.''  Dokl. Akad. Nauk SSSR {\bf 30},301-305(1941).

[14] F. Bonetto, J. Lebowitz, and L. Rey-Bellet ``Fourier's law: a challenge for theorists.'' pp. 128-150 in {\it Mathematical Physics 2000}, A. Fokas, A. Grigoryan, T. Kibble and B. Zegarlinsky (eds), Imperial College, London, 2000.

[15] D. Dolgopyat and C. Liverani ``Energy transfer in a fast-slow Hamiltonian system.''  Commun. Math. Phys. {\bf 308},201-225(2011).

[16] D. Ruelle ``A mechanical model for Fourier's law of heat conduction.''  Commun. Math. Phys. {\bf 311},755-768(2012).

[17] L. Bertini, A. De Sole, D. Gabrielli, G. Jona-Lasinio, and C. Landim ``Towards a nonequilibrium thermodynamics: a self-contained macroscopic description of driven diffusive systems.''  J. Statist. Phys. {\bf 135},857-872(2009).

[18] B. Derrida, J.L. Lebowitz, and E.R. Speer ``Exact free energy functional for a driven diffusive open stationary nonequilibrium system.''  Phys. Rev. Lett. {\bf 89},030601 (2002).

[19] F. Anselmet, Y. Gagne, E.J. Hopfinger, and R.A. Antonia ``High-order velocity structure functions in turbulent shear flows.''  J. Fluid Mech. {\bf 140},63-89(1984).

[20] A. Vincent and M. Meneguzzi ``The spatial structure and statistical properties of homogeneous turbulence.''  J. Fluid Mech. {\bf 225},1-20(1991).

\end